\documentclass{article}
\usepackage{spconf,amsmath,graphicx}
\usepackage{multirow}
\usepackage{booktabs}
\usepackage{amssymb}

\usepackage[table]{xcolor} 
\usepackage{enumitem}
\setlist{nosep, leftmargin=14pt}
\usepackage{booktabs}
\usepackage{url}
\usepackage{soul}
\usepackage[normalem]{ulem}
\title{AGE2HIE: Transfer Learning from Brain
Age to Predicting Neurocognitive
Outcome for Infant Brain Injury}

\name{Rina Bao$^*$, Sheng He$^*$\thanks{$^*$ equal contribution}
, Ellen Grant, Yangming Ou}
\address{Boston Children's Hospital, Harvard Medical School}

\begin{document}
%\ninept
%

\maketitle
%
%\begingroup\renewcommand\thefootnote{\textsection}
%\footnotetext{Equal contribution}

\begin{abstract}
Hypoxic-Ischemic Encephalopathy (HIE) affects 1 to 5 out of every 1,000 newborns, with 30\% to 50\% of cases resulting in adverse neurocognitive outcomes. However, these outcomes can only be reliably assessed as early as age 2. Therefore, early and accurate prediction of HIE-related neurocognitive outcomes {\color{black}using deep learning models} is critical for improving clinical decision-making, guiding treatment decisions and assessing novel therapies. 
However, a major challenge in developing deep learning models for this purpose is the scarcity of large, annotated HIE datasets. We have assembled the first and largest public dataset, however it contains only 156 cases with 2-year neurocognitive outcome labels. In contrast, we have collected 8,859 normal brain black Magnetic Resonance Imagings (MRIs) with 0-97 years of age that are available for brain age estimation using deep learning models.
In this paper, we introduce AGE2HIE to transfer knowledge learned by deep learning models from healthy controls brain MRIs to a diseased cohort, from structural to diffusion MRIs, from regression of continuous age {\color{black}estimation} to prediction of {\color{black}the} binary neurocognitive outcomes, and from lifespan age (0-97 years) to infant (0-2 weeks). Compared to
training from scratch, transfer learning from brain age estimation significantly improves not {\color{black}only the} prediction accuracy (3\% or 2\% improvement in same or multi-site), but also {\color{black}the} model generalization {\color{black}across different sites} (5\% improvement in cross-site validation).

\end{abstract}
\section{introduction}

%Reducing mortality and morbidity associated with HIE is, therefore, a critical public health objective.  

%There is a pressing need to prediction neurocognitive outcomes, The early identification of patients at high risk for adverse outcomes sooner after therapy initiation remains a significant challenge, as clinical outcomes are often not reliably measurable until the age of two years~\cite{laptook2021limitations}. This highlights the critical need for accurate, reliable early prognosis and precise outcome prediction. The accurate identification high risk of adverse outcomes with neonatal brain magnetic resonance images (MRIs) are critical steps toward this goal, as they enable the early identification of high-risk patients, a better understanding of neurological symptoms, and the early evaluation of treatment efficacy. 

Neonatal hypoxic-ischemic encephalopathy (HIE) remains a significant public health concern, characterized by brain injury due to insufficient blood and oxygen supply to the brain. Despite the adoption of Therapeutic Hypothermia (TH) as the standard of care, 30\%-50\% of HIE patients {\color{black}experience} adverse neurocognitive outcomes~\cite{shankaran2005whole,edwards2010neurological}. Therapeutic innovation is slow and inconclusive since patients at high risk of developing adverse outcomes cannot be accurately identified before therapy and outcomes cannot be reliably measured after therapies until age 2 years~\cite{laptook2021limitations}. Therefore, there is a pressing need to develop algorithms for predicting 2-year neurocognitive outcomes.
% as current therapeutic innovation is hindered by the inability to identify high-risk patients before therapy and the delayed measurement of outcomes until two years after therapy~\cite{laptook2021limitations}, highlighting the lack of neonatal biomarkers that could guide targeted interventions, improve trial efficiency, and accelerate therapeutic advancements.

The development and efficiency of {\color{black}deep learning} technologies rely heavily on access to extensive {\color{black}annotated} datasets. Our lab's decade-long effort produced the BONBID-HIE dataset~\cite{bao2023boston,bao2024bostonneonatalbraininjury}, which, despite being the largest publicly available HIE dataset (N=156 cases with 2-year neurocognitive outcomes), falls short of the thousands of samples typically required to achieve optimal performance {\color{black} of deep learning models}~\cite{deng2009imagenet}. Fortunately, leveraging data from non-target cohorts {\color{black}(known as Transfer learning~\cite{neyshabur2020being})} has proven effective in enhancing performance {\color{black}of deep learning models} across various clinical prediction tasks.  {\color{black} The deep learning models pre-trained} using millions of natural images or thousands of unrelated medical images have shown to improve accuracy in multiple target conditions {\color{black}with fine-tunning on the target datasets}, such as brain tumor classification~\cite{deepak2019brain}, stroke lesion subtyping~\cite{jung2020deep}, and abnormality detection~\cite{talo2019application}.

% why not transfer from other models? ImageNet~\cite{deng2009imagenet} has generated many benchmark models. Many of these ImageNet-based benchmark
% models have been transferred to brain MRIs ~\cite{raghu2019transfusion,shin2016deep,tajbakhsh2016convolutional} and why not doing the same for HIE outcome predictions on brain MRIs? utilizing brain age prediction as a pre-training task leverages the availability of large-scale, objective data, enabling the model to learn generalized MRI features that can be transferred to more specific tasks like HIE outcome prediction. This approach provides a robust foundation for overcoming the limitations of small sample sizes in disease-specific datasets, improving model performance through transfer learning.

\begin{figure*}[!t]
    \centering
\includegraphics[width=0.9\linewidth]{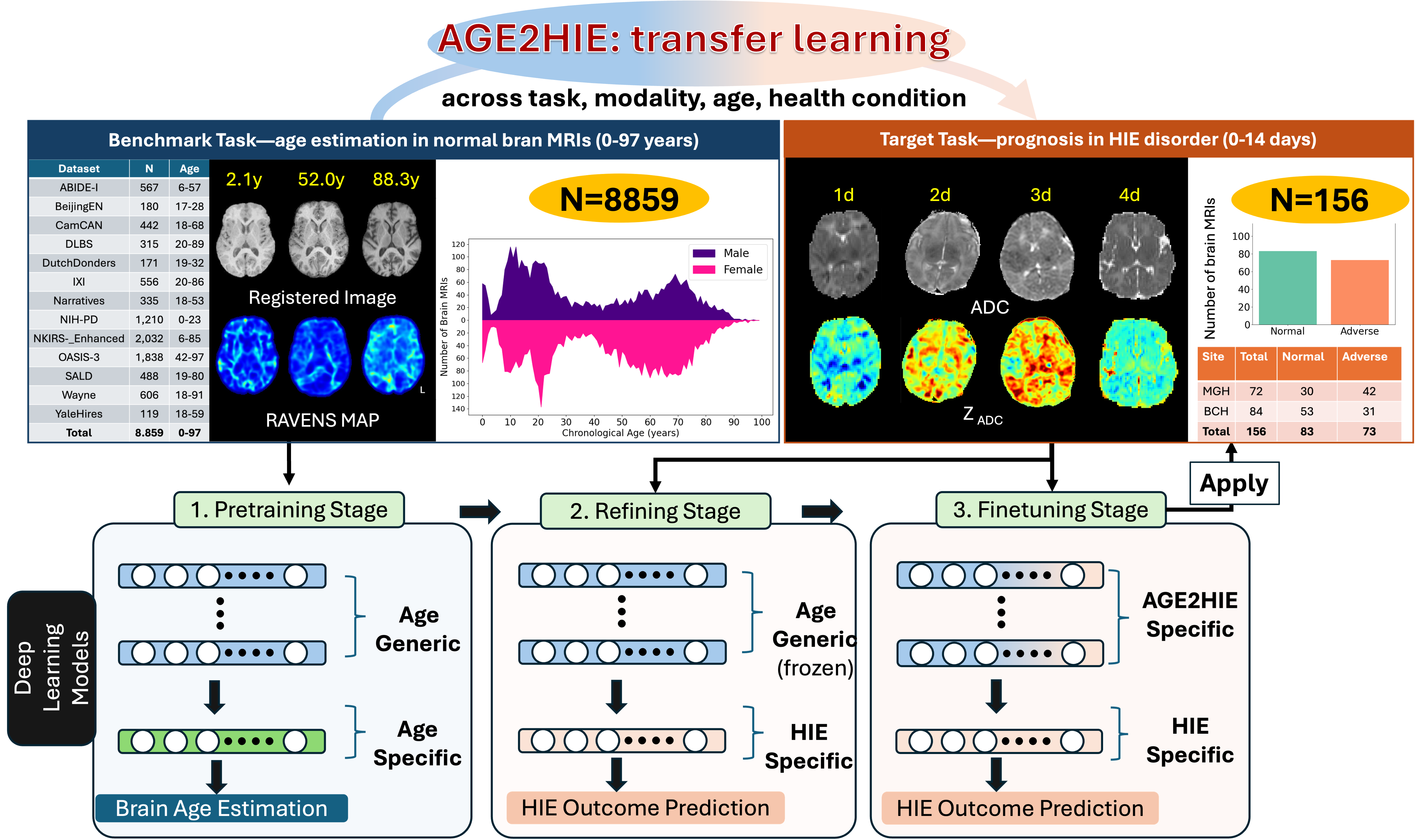}
    \vspace{-0.4cm}
    \caption{{The framework of the proposed AGE2HIE for transferring knowledge learned by deep learning models from brain age estimation (benchmark task, left-upper pannel, with exmaple images and data distribution) with 8,859 samples to HIE outcome prediction (target task, right-upper pannel, with exmaple images and data distribution) with only 156 subjects. The bottom row shows the three stages of transfer learning: 1, Pretraining Stage trains all layers of the model which contains the age generic layers (blue color) and age specific layer (green color); 2, Refining Stage replace the age specific layer to HIE specific layer (orange color) and only trains this HIE specific layer, and 3, Finetuning Stage fine-tunes all layers of the deep learning model to learn the AGE2HIE Specific layers which contains age generic and HIE specific information.}}
    \label{fig:workflow}
    \vspace{-0.5cm}
\end{figure*}

Although transfer learning from {\color{black}natural images on} ImageNet~\cite{deng2009imagenet} {\color{black}to medical images} has been widely used for {\color{black}tasks on brain MRIs}~\cite{shin2016deep,tajbakhsh2016convolutional}, {\color{black}deep learning models pre-trained on ImageNet} are designed for natural images in 2D. {\color{black}When transferring these models to 3D brain MRIs,} {\color{black}they} process the {\color{black}3D} MRIs slice by slice, which may not fully exploit the 3D nature of the {\color{black}MRI volumes}, especially when the infant brain MRI diffusion images have 2-4mm slice thickness.

In this {\color{black}paper}, we introduce AGE2HIE, a novel study design that leverages pre-trained {\color{black}3D deep learning} models from brain age estimation {\color{black}task on a large dataset of 8,859 brain MRI volumetric T1-weighted scans} to HIE outcomes {\color{black}prediction on a small dataset of 156 infant brain diffusion MRIs}, addressing this small sample size problem in HIE outcome prediction task.
%\st{Deep learning has been used for lesion detection~\cite{murphy2017automatic}{\color{black} in brain MRIs}, but not {\color{black}for} outcome prediction directly in brain MRI of HIE patients.}
Our {\color{black}proposed AGE2HIE} transfers a 3D {\color{black}deep learning} model (a) across tasks: from estimating the continuously-valued ages {\color{black}in healthy brain MRIs} to predicting the binary neurocognitive outcomes in HIE patients; (b) across age: from 0-97 years to infants 0-2 weeks; (c) across MRI modalities: from 3D volumetric T1-weighted MRI where age estimation {\color{black}deep learning} model was developed to 3D diffusion MRI of HIE patients; and (d) across health conditions: from normal controls for age estimation to patients with HIE abnormalities for outcome prediction.

%In summary, this paper presents three key contributions. First, it is the initial study to apply deep learning for predicting 2-year neurocognitive outcomes in HIE. Second, we introduce a novel pipeline, Age2HIE, which leverages large-scale brain MRI data for age prediction as a strategy to address the challenge of limited sample sizes in HIE outcome prediction. Finally, our experiments demonstrate the critical importance of pretraining on brain age, highlighting its effectiveness and generalization for HIE outcome prediction across various scenarios, including multi-site and cross-site datasets.

\section{Methodology}
{\color{black}This section presents the proposed AGE2HIE framework, including two main parts (three stages,as shown in Fig.~\ref{fig:workflow}) : pre-training the deep learning model (Pretraining stage) for brain age estimation (benchmark task)  and transferring the pre-trained deep learning model (Refining Stage and Finetuning Stage) for HIE outcome prediction (target task).}

\subsection{Pre-training for brain age {\color{black}estimation}}
%Using brain age as a pretraining task as it offers a large-scale, objective groudtrtuh compared to many other disease-specific domain tasks. Brain age estimation leverages an abundance of available data easier, allowing the model to learn features that are not only generalizable but also rich in developmental patterns. This contrasts with many disease-specific datasets, which tend to be smaller and more subjective in nature.
\textbf{Rationale of using brain age {\color{black}estimation} as the benchmark task.} We chose brain age estimation as the benchmark task for two reasons. {\color{black}The age range is} in 0-97 years, offering the diversity to learn rich and generalizable patterns in 3D MRI{\color{black}s}. Diversity in the benchmark task is important, as 2D imageNet saturated its accuracy and {\color{black}maximized} its accuracy in transfer{\color{black}ring} when over 1000 classes were available in the benchmark training~\cite{t15huh2016makes}. A second reason is that age is available for all the subjects with 3D brain MRI{\color{black}s}, and is {\color{black}a reliable label given the birth and MRI scan date of the subject}, where as other properties such as disease diagnosis or lesion annotations are not available for everyone or they are not as reliable as age for a ground truth. 
%In realistic scenarios of small sample sizes, AI can rely on big data from other tasks to help address the small sample problem in target diseases. AI models are designed with a plethora of parameters to predict specific outcomes. Notably, parameters proximal to the input MRI images encapsulate generalized information pertinent to MR images, which is applicable across various tasks. Conversely, parameters nearer to the output layer harbor task-specific details~\cite{yosinski2014transferable}. This hierarchical structuring of information allows for a segment of the parameters, particularly those capturing generic imaging features, to be pre-trained on auxiliary tasks using expansive datasets, subsequently being repurposed for predicting outcomes in HIE. Furthermore, the layers of a deep learning model closer to the input MRI capture generic information about MRI characteristics, such as tissue intensity patterns and anatomical structures, which can be beneficial across various tasks. By training these initial layers on a large-scale brain age dataset, we can effectively reuse these generalized MRI features for HIE outcome prediction, providing a strong foundation through transfer learning(see Figure~\ref{fig:workflow}). This approach allows the model to leverage pre-existing knowledge in brain MRIs, thereby improving performance compared to training from scratch on smaller, task-specific datasets.

\textbf{Brain MRI{\color{black}s} {\color{black}for} Age Estimation.} We downloaded and merged 13 public datasets, resulting in a total of 8,859 normal brain MRIs spanning 0-97 years of age~\cite{he2023human}. This dataset serves as our 3D large-scale brain MRI datasets, which we used to pre-train the {\color{black}deep learning} model on 3D MRIs. We noted that the only consistent information across all MRI scans in the 13 datasets was the age and sex of the subjects. Therefore, brain age {\color{black}estimation} was selected as the {\color{black}benchmark task}. The 13 public datasets~\cite{he2023human} {\color{black}include} ABIDE-I~\cite{di2014autism}, BeijingEN~\footnote{\url{http://fcon 1000.projects.nitrc.org/index.html}}, CamCAN~\cite{taylor2017cambridge}, DLBS~\cite{park2012neural},
DutchDonders~\footnote{\url{ https://data.donders.ru.nl/?1}},
IXI~\footnote{\url{ https://brain-development.org/ixi-dataset}},
Narratives~\cite{nastase2021narratives},
NIH-PD~\cite{evans2006nih},
NKI-RS\_Enhanced~\cite{nooner2012nki},
OASIS-3~\cite{lamontagne2018oasis},
SALD~\cite{wei2017structural},
Wayne~\cite{daugherty2017virtual},
YaleHires~\cite{finn2015functional}.
{\color{black}The upper left box in
Fig.~\ref{fig:workflow} shows the sample size and age
ranges in each dataset, as well as the distribution of ages in
the merged database.} 
%For details of these 13 datasets, please refer to \cite{he2023human}.
These T1-weighted brain MRIs were all acquired at 1 isotropic mm voxel size or slightly higher resolution.
The pre-processing of {\color{black}T1-weighted} images is same as {\color{black} in other studies}~\cite{he2023human,he2022deep}, including N4 bias correction, field of view normalization,  Multi-Atlas Skull Stripping (MASS) and deformable registration to Atlas.
{\color{black} The  non-rigidly registration method is used to register each T1-weighted MRI to an Atlas, which can generate two different channels~\cite{he2021multi}: registered image representing the contrast information and RAVENS map quantifing voxel-wise morphological information (as shown in Fig.~\ref{fig:workflow}). }

\textbf{Pre-training for Brain Age Estimation.} 
%Transfer learning for 2D medical image tasks are benefit from pretraining~\cite{raghu2019transfusion,shin2016deep,tajbakhsh2016convolutional} on 2D ImageNet~\cite{deng2009imagenet}. Therefore, transfer learning for 3D MRI studies requires a large-scale 3D ImageNet. 
The deep learning model ResNet18-3D~\cite{he2016deep} {\color{black}(which is the best model for HIE outcome prediction, see Table~\ref{table:performance} in Results section for justification)} was pre-trained for brain age estimation, using mean absolute error (MAE) as the loss function, which is commonly used in brain age estimation studies~\cite{he2022deep,he2023human}. During the pretraining stage, we utilized the Adam optimizer~\cite{kingma2014adam} with an initial learning rate of 0.001, which was halved every 20 epochs, over a total of 80 training epochs. The model was trained with a batch size of 16, following the default parameters provided by the PyTorch platform, consistent with previous works~\cite{he2022deep,he2023human}.

\begin{table*}[htb]
\centering
\caption{Accuracy {without ($\times$) vs with ($\checkmark$)  transferring} age-pretrained model to HIE outcome prediction.} 
\resizebox{0.85\linewidth}{!}{
\begin{tabular}{cc|l|c|l|l|l}
\toprule
 \parbox[t]{1mm}{\multirow{7}{*}{\rotatebox[origin=c]{90}{Train/Test}}}  & \parbox[t]{5mm}{\multirow{7}{*}{\rotatebox[origin=c]{90}{within the same site}}} & \bf{Setting}&\textbf{Transfer} & \textbf{Accuracy (\%)}  & \textbf{Sensitivity (\%)}  & \textbf{Specificity (\%)}  \\ 
\cmidrule{3-7}

&&{Training on MGH (N=72)} &  {$\times$}                   & 72.28$\pm$17.92           & 70.00$\pm$29.15              & 71.21$\pm$18.26              \\
&&Testing on MGH& \checkmark                   & \cellcolor{gray!30}74.85$\pm$9.21            & \cellcolor{gray!30}74.76$\pm$21.27              & \cellcolor{gray!30}75.50$\pm$17.51              \\ 
\cmidrule{3-7}
&&{Training on BCH (N=84)} &  {$\times$}       &   51.24$\pm$8.47                &            58.29$\pm$18.35       &       41.33$\pm$18.73                     \\
&&Testing on {BCH} & \checkmark                     &      \cellcolor{gray!30}54.38$\pm$12.19               &       \cellcolor{gray!30}66.92$\pm$23.46              & \cellcolor{gray!30}46.05$\pm$20.07    \\
\cmidrule{3-7}
&&{Training on MGH+BCH (N=156)}& {$\times$}  &68.66$\pm$8.5 &78.82$\pm$12.95 &66.52$\pm$16.27 \\
&&Testing on MGH+BCH &\checkmark& \cellcolor{gray!30}70.45$\pm$9.1& \cellcolor{gray!30}77.35$\pm$14.28 & \cellcolor{gray!30}66.83$\pm$15.16\\ 
 \bottomrule
\end{tabular}
}
\vspace{-0.5cm}
\label{table:all}
\end{table*}

\begin{table*}[htb]
\centering
\caption{{\color{black}Generality without ($\times$) vs with ($\checkmark$)  transferring age-pretrained model to HIE outcome prediction}. }
\resizebox{0.8\linewidth}{!}{%

\begin{tabular}{cc|l|c|l|l|l}
\toprule
\parbox[t]{1mm}{\multirow{5}{*}{\rotatebox[origin=c]{90}{Train/Test}}}  & \parbox[t]{5mm}{\multirow{5}{*}{\rotatebox[origin=c]{90}{cross-site}}} 
& \textbf{Setting}        & \textbf{Transfer}   & \textbf{Accuracy (\%)}  & \textbf{Sensitivity(\%)}  & \textbf{Specificity(\%)}  \\ 
\cmidrule{3-7}
&& 
 {\color{black}Training in MGH (N=72)}&      {\color{black}$\times$}                & 57.14$\pm$18.99           & 90.00$\pm$20.00              & 49.36$\pm$19.78 \\ 
  && {\color{black}Testing in BCH (N=84)} &  \checkmark                   & \cellcolor{gray!30}62.57$\pm$9.79            &\cellcolor{gray!30}84.29$\pm$13.93              & \cellcolor{gray!30}52.21$\pm$17.56 \\ 
\cmidrule{3-7}
 && {\color{black}Training in BCH (N=84)}&{\color{black}$\times$} &    72.50$\pm$11.27   &  60.00$\pm$13.33  & 75.88$\pm$18.69 \\
 && {\color{black}Testing in MGH (N=72)} &\checkmark    & \cellcolor{gray!30}77.21$\pm$8.55                        &          \cellcolor{gray!30}64.43$\pm$11.27                 &   \cellcolor{gray!30}82.57$\pm$14.92                      \\
 \bottomrule
\end{tabular}
}
\label{table:split}
\vspace{-0.5cm}
\end{table*}

\subsection{HIE Outcome Prediction with Transfer Learning}

% The input to the deep learning networks for predicting 2-year neurocognitive outcomes consists of Apparent Diffusion Coefficient (ADC) maps and $Z_{ADC}$. The output is a binary classification: adverse outcome (label 1) or normal outcome (label 0). Thus, this represents a binary classification problem.

\textbf{BONBID-HIE MRI with neurocognitive outcome.} The first public and largest HIE dataset--BONBID-HIE~\cite{bao2023boston,bao2024bostonneonatalbraininjury} provided 156 cases from a cohort of neonates diagnosed with HIE at Massachusetts General Hospital (MGH) and Boston Chilren's Hospital (BCH). Details of the dataset are shown in the upper right panel of Fig.~\ref{fig:workflow}. For each case, the provided inputs included Apparent Diffusion Coefficient (ADC) maps and $Z_{ADC}$, which were used as inputs for the algorithm containers. The output of algorithm was 2-year neurocognitive outcome, which is binary (normal vs abnormal, and roughly 1:1 ration in our data.). MRIs were acquired on either GE 1.5T Signa scanner or Siemens 3T Trio scanner. $Z_{ADC}$ maps are designed to normalize and make ADC values comparable across brain voxel locations~\cite{bao2023boston} by quantifying location-specific deviations from normal. %which is important for abnormal regions on brain MRIs. $Z_{ADC}$ maps quantify location-specific deviations from normal. 
%(1) A normative ADC atlas was generated from scans of 13 neonates. This atlas quantifies the mean ADC values and standard deviations at every voxel~\cite{50ou2017using}. (2) A deformation field $D$ is computed, which maps each voxel $x$ in the patient's ADC map to its anatomically corresponding location $D(x)$ in the atlas space. The normal range of ADC variation is defined by the mean $\mu D(x)$ and standard deviation $\sigma D(x)$ for  each voxel. (3) The patient's ADC value $I_x$ at voxel $x$ is converted to a $Z$-score: $Z_{ADC}(x) = \frac{I_x - \mu D(x)}{\sigma D(x)}$. Therefore, 
%The $Z_{ADC}$ value at voxel location $x$ quantifies how many standard deviations the patient’s ADC value $I_x$ deviates from the mean normal ADC value at that anatomical location. For details on the data, please refer to~\cite{bao2023boston}. Data of neurocognitive outcomes around 2 years of age. Outcome is defined as normal versus adverse, according to clinical criteria and NRN recommendations~\cite{laptook2017effect,shankaran2017effect}. 
% \begin{table}[h]
% \centering
% \caption{Data of HIE 2-year outcomes with MRIs.}
% \begin{tabular}{c |c| c| c}
% \toprule
% Site&N & normal outcome & adverse outcome\\
% \midrule
% BCH & 72  &30 &42\\ 
% MGH & 84 & 53 & 31\\
% \midrule
% Total& 156 & 83 & 73\\
% \bottomrule
% \end{tabular}\label{tab:data}
% \end{table}

{\color{black}The transfer learning part of our} proposed {\color{black}AGE2HIE} pipeline for HIE outcome prediction consists of two stages to transfer {\color{black}the knowledge learned by deep learning models} from brain age estimation, as illustrated in Fig.~\ref{fig:workflow}: {\color{black} refining  stage and fine-tuning stage}. This two-stage process aims at reducing overfitting by gradually tailoring the age estimator in a controlled manner to the specific HIE domain.

{\color{black}\textbf{Refining Stage: Refining the last layer}. In the refining stage, we removed the last fully connected layer (with one output) of ResNet18-3D for brain age estimation and replaced a new fully connected layer with randomly initialized parameters (with two outputs). We fixed the parameters on all convolutional layers for feature extraction. We refined this new last fully connected layer for HIE outcome prediction with cross entropy loss. The last layer is refined with 100 epochs with a learning rate of 0.001. }

{\color{black}\textbf{Fine-tuning Stage: Fine-tuning the entire network.} After the refinement of the last layer, all parameters in ResNet18-3D were fine-tuned for HIE outcome prediction with a low learning rage 0.0005 for another 100 epochs.}

{\color{black}\subsection{Experimental setting}}

{\color{black}We used 5-fold cross-validation to evaluate the performance. We randomly split the 156 subjects into 5 folds approximate equally without overlap.
Each time, we took one fold out for testing and the rest is used for training.
The average accuracy and standard deviation over 5-fold cross validation is reported in the following sections. 
To evaluate the \textbf{accuracy} of the transfer learning with Train/Test within the same site: we fine-tune  the model on all 156 subjects with 5-fold cross-validation.
We also fine-tune and test the accuracy on each site (only on MGH and only on BCH) with 5-fold cross-validation.
To evaluate the  \textbf{generality} of the transfer learning with Train/Test cross-site: we fine-tune  the model on one site (MGH or BCH) and test it on another site (BCH or MGH).
The accuracy is measured with three metrics: accuracy, sensitivity and specificity.}

\section{Results}

{\color{black}
%\textbf{Accuracy of transfer learning with Train/Test with the same site}: 
\textbf{AGE2HIE with transfer learning improves accuracy}: Table~\ref{table:all} shows the accuracy with and without transfer learning with Train/Test with the same site.
Transfer
learning improved the average accuracy by 2.57\%, 3.14\%, and 1.79\% in MGH data (N=72), BCH data(N=84) and
MGH+BCH data (N=156), respectively.
The same trend is also found in the average sensitivity and specificity (expect the sensitivity on MGH+BCH data).
%(2) the average sensitivity by 4.76\%, 8.63\%, and -1.47\%; (3) the average specificity by 4.29\%, 4.72\%, and 0.31\%
The large improvement is obtained in BCH site where the accuracy, sensitivity and specificity are lower than MGH and combeined (MGH+BCH).
The results demonstrate that transfer learning can improve the accuracy on the site with low accuracy without transfer learning. 
%Another observation is that the accuracy in BCH is lower than it in MGH with/without transfer learning. 
%The main reason is that the MGH cohort and the BCH cohort differ greatly in ?????
}

{\color{black}
%\textbf{Generality of transfer learning with Train/Test cross-sites}:
\textbf{AGE2HIE with transfer learning enhances generality}:
Table~\ref{table:split} shows the generality with and with transfer learning with training on one site and testing on another site.
Transfer
learning improved the average accuracy by 5.43\% when training on MGH and testing on BCH and 4.71\% when training on BCH and testing on MGH.
The same trend is also found in the average sensitivity and specificity (expect the sensitivity when training on MGH and testing on BCH).
The results indicate that transfer learning can also help to improve the generality of the deep learning model.
}

\textbf{Comparison with existing works.} To date, there is a lack of MRI-based machine learning studies focused on HIE. Most existing studies have been conducted by clinical teams using statistical analyses to demonstrate correlations between MRI patterns and neurocognitive outcomes in HIE~\cite{trivedi2017validated}. However, these studies primarily rely on private datasets with limited sample sizes. For example, study by Trivedi. et al.~\cite{trivedi2017validated} reported a sensitivity of 0.77 and a specificity of 0.46, which are lower than the performance achieved in our multi-site evaluation {\color{black}(sensitivity of 0.7882 and specificity of 0.6683)}. 

{\color{black}\textbf{Ablation Study for different ResNet-3D.}} We conducted an ablation study to evaluate the performance of {\color{black}3D ResNet with different number of layers} as the backbone for HIE outcome prediction. As shown in Table \ref{table:performance},
 ResNet18-3D {\color{black} with 18 layers} was chosen as the backbone for {\color{black}transfer learning} due to its balance between model complexity and performance. While ResNet50 {\color{black} with 50 layers} is deeper {\color{black}than ResNet18}, {\color{black} which usually achieves better results than ResNet18 on large datasets, gives lower accuracy than ResNet18 on 3D brain MRIs with only 156 samples for HIE outcome prediction.} 
{\color{black} The ResNet18-3D has small number of layers,} reduces overfitting and performs robust feature extraction, which is essential for tasks with limited data. Therefore, ResNet18-3D was selected as the preferred backbone for our HIE outcome prediction, offering a balance between performance, model complexity, and computational efficiency.

\begin{table}[!tb]
\centering
\caption{{\color{black}Comparison of ResNet-$x$ 3D with different number of layers $x$} without transfer learning on HIE outcome prediction.}
\resizebox{0.7\linewidth}{!}{%
\begin{tabular}{c|c|c}
\toprule
\textbf{Model} & \#\textbf{Layer} & \textbf{Accuracy} \\
\midrule
\rowcolor{gray!30} 
Resnet18-3D &18 &  68.66\% $\pm$ 8.5\% \\
Resnet34-3D & 34 & 57.77\% $\pm$ 8.2\% \\
Resnet50-3D & 50 & 66.70\% $\pm$ 7.4\% \\
\bottomrule
\end{tabular}
}
\label{table:performance}
\vspace{-0.5cm}
\end{table}

\section{Discussion and Conclusion}
In this {\color{black}paper}, we introduced AGE2HIE, a novel transfer learning approach that leverages knowledge {\color{black}learned by deep learning models} from brain age estimation {\color{black}task} to HIE outcome prediction. 
{\color{black}We evaluated the proposed AGE2HIE on BONBID-HIE datasets with 156 subjects. Our experimental results show the advantages of transfer learning for brain age estimation to HIE outcome prediction with small
samples.
}
{\color{black}The proposed AGE2HIE} consistently boosts the performance of deep learning models for both accuracy and generality, effectively addressing the limitations of the small HIE dataset. In future work, we plan to more thoroughly test how the effect of transfer learning is impacted
by the sample sizes and  larger range of MRI timepoints in the benchmark {\color{black}of} brain age estimation, by the sample size in the target cohort, and by different benchmark tasks (e.g., 2D ImageNet-based or 3D MRI-based tasks).
\section*{Acknowledgement}

This work was funded, partly, by Harvard Medical
School and Boston Children’s Hospital via Early Career Development Award, Thrasher Research Fund Early Career Awards 02402, NIH R21NS121735, R61NS126792, and R03HD104891.

{\small
    \bibliographystyle{IEEEbib}
    \bibliography{refs}
}

%\setlength{\bibitemsep}{1pt}
%\printbibliography

\end{document}